\documentclass[aps,pre,twocolumn,showpacs,superscriptaddress]{revtex4}
\usepackage{graphicx,epsfig}
\begin{document}
\title{Phase growth in bistable systems with impurities}
\author{C. Echeverria} 
\affiliation{Laboratorio de F\'isica Aplicada y Computacional, Universidad Nacional Experimental del T\'achira, San Crist\'obal, Venezuela.}
\affiliation{Centro de F\'isica Fundamental, Universidad de Los
Andes, M\'erida, M\'erida 5251, Venezuela.}
\author{K. Tucci}
\affiliation{Centro de F\'isica Fundamental, Universidad de Los
Andes, M\'erida, M\'erida 5251, Venezuela.}
\affiliation{SUMA-CeSiMo, Universidad de Los Andes, M\'erida, M\'erida 5251, Venezuela.}
\author{M. G. Cosenza}
\affiliation{Centro de F\'isica Fundamental, Universidad de Los
Andes, M\'erida, M\'erida 5251, Venezuela.}

\date{\today}

\begin{abstract}
A system of coupled chaotic bistable maps on a lattice with randomly distributed impurities is investigated as a model
for studying  the phenomenon of phase growth in nonuniform media. The statistical properties of the system are characterized by means of the average size of spatial domains of equivalent spin variables
that define the phases. It is found that the rate at which phase domains grow becomes smaller when impurities are present and that the average size of the resulting domains in the inhomogeneous state of the system decreases when the density of impurities is increased. The phase diagram showing regions where homogeneous, heterogeneous, and chessboard patterns occur on the space of parameters of the system is obtained. A critical boundary that separates the regime of slow growth of domains from the regime of fast growth in the heterogeneous region of the phase diagram is calculated. The transition between these two growth regimes is explained in terms of the stability properties of the local phase configurations. 
Our results show that the inclusion of spatial inhomogeneities can be used as a control mechanism for the size and growth velocity of phase domains forming in spatiotemporal systems.
\end{abstract}

\maketitle

\section{Introduction}
There has been much recent interest in the study of spatiotemporal
dynamical processes on nonuniform or complex networks. In this
context, coupled map lattices \cite{Kaneko} have provided fruitful
and computationally efficient models for the investigation of a
variety of dynamical processes in spatially distributed systems.
In particular, the discrete-space character of coupled map systems
makes them specially appropriate for the investigation of
spatiotemporal dynamics on nonuniform networks that can represent
models of heterogeneous media. 
The nonuniformity may be due to the intrinsic heterogeneous nature of
the substratum, typical of pattern formation in biological
contexts, or it may arise from random imperfections or
fluctuations in the medium at some length scales. Such heterogeneities can have
significant effects on the formation of spatial patterns, for example,
they can induce reberverators in excitable media and defects can
serve as nucleation sites for domain growth processes.

Recently, the study of
the phase-ordering properties of systems of coupled chaotic maps
and their relationship with Ising models in statistical physics has been a focus of
attention \cite{Chate,Chate2,Wei,Stra,Just,A1,TCA1}. These works have mainly assumed
the phase competition dynamics taking place on an ordered spatial support. 
This article investigates the phenomenon of phase growth in 
coupled chaotic maps on lattices with randomly distributed impurities as a model for studying this 
process on nonuniform media. In particular, this model of coupled maps on nonuniform networks yields a scenario
to explore the role that the local configurations of the underlying lattice play on the statistical properties of phase ordering processes on spatiotemporal systems.
In addition, this model can be seen as a simpler but computationally more efficient alternative
to the study of phase separation phenomena on media with impurities than conventional computational fluid dynamics techniques. 

In Sec.~II, we present a procedure for the construction of a lattice with impurities and define the coupled map model
on this network. The phase growth dynamics in the presence of impurities is studied in Sec.~III. The phase diagram of the
system on the space of its parameters is obtained in this Section. Section~IV contains the conclusions of this work.

\section{Coupled map lattice with impurities}
To generate a lattice with impurities, we start from a two-dimensional array of cells of size $L \times L$ with periodic boundary conditions and remove a given fraction $\rho$ of cells at random. The removed sites can be considered as  impurities,  defects, or random imperfections on the spatial support. We define an impurity as a non-active site, i.e., a site that possesses no dynamics. The fraction of impurities in the lattice can be characterized in terms of the minimum Euclidean distance $d$ between impurities, as shown in Figure~1. In this way, the density $\rho$ scales with the minimum distance between impurities as $\rho=0.625 d^{-2}$.  We shall use values $d \geq 2$ in order to avoid contiguous impurities on the lattice.
\begin{figure}[ht]
\centerline{\includegraphics[width=0.4\linewidth,angle=90]{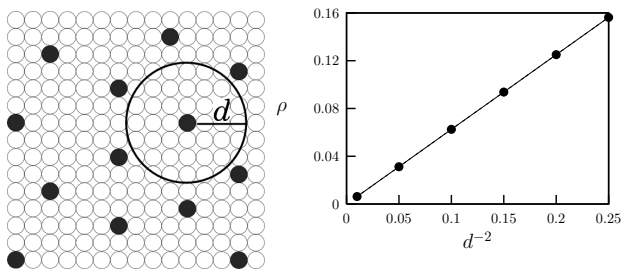}}
\caption{\label{FIG1}(Left) Spatial support showing active sites ($\circ$) and impurities ($\bullet$). 
Impurities are randomly placed keeping a minimum distance $d$ between them. (Right) The density of impurities scales as $\rho=0.625 d^{-2}$.}
\end{figure}

The $N=L^2(1-\rho)$ active sites can be enumerated by an index $i=1,\ldots,N$.
The equations describing the
dynamics of the diffusively coupled map system defined on such
nonuniform lattice are
\begin{equation}
\label{CML}
x^i_{t+1} = (1-\epsilon)f(x^i_t) + \frac{\epsilon}{{\cal{N}}^i} \sum_{j\in {\nu^i}} f(x^j_t)\;,
\end{equation}
where $x^i_t$ gives the state of an active site $i$  at time $t$; $\nu^i$
is the set of the nearest active neighbors (von Neumann neighborhood) of site $i$, and ${\cal{N}}^i \in \{1,2,3,4\}$ is
the cardinality of $\nu^i$; the parameter $\epsilon \in [0,1]$ measures the coupling strength, and $f(x)$ is a map that expresses the local dynamics. Impurities are not coupled to any other cell in the lattice. 

The above coupled map equations on a lattice with impurities can be generalized to include other coupling schemes and neighborhoods, higher dimensions or continuous-time local dynamics. Different spatiotemporal phenomena can be also be studied on such nonuniform structures by providing appropriate local dynamics and couplings.

In order to describe a bistable dynamics, we assume a piecewise
local map \cite{MH1}
\begin{equation}
f(x) = \left\lbrace
\begin{array}{ccl}
-2\mu/3-\mu x, & \mbox{if} & x \in [-1,-1/3] \\
\mu x, & \mbox{if} & x \in [-1/3,1/3] \\
2\mu/3 - \mu x, & \mbox{if} & x \in [1/3,1] \;.
\end{array}\right. 
\end{equation} 
When the parameter $\mu \in (1,2)$ the map has two symmetric chaotic band attractors, 
one with values 
$x^i_t >0$ and the other with $x^i_t<0$, separated
by a finite gap about the origen. Then the local states have two well defined 
symmetric phases that can be characterized by spin variables
defined as the sign of the state at time $t$, $\sigma^i_t = \mbox{sign}(x^i_t)$.

\section{Phase growth in the presence of impurities}
We fix the local parameter at $\mu=1.9$ and set the initial conditions as follows: 
one half of the active sites are randomly
chosen and assigned random values uniformly distributed
on the positive attractor while the other half are similarly 
assigned values on the negative attractor. If the number of active sites
is odd, then the state of the remaining site is
assigned at random on either attractor.

In regular lattices ($\rho=0$) phase growth occurs for values  $\epsilon > \epsilon_o$, where $\epsilon_o = 0.67$ \cite{Chate}. In contrast, in the medium with impurities there exist a minimum value of $\rho$ for which the domains formed by the two phases reach a frozen configuration for all values of the coupling $\epsilon$.
Figure~2 shows stationary patterns emerging in the system for different values of parameters. The top panels
show that the average size of domains decreases when the density of impurities in the system is increased. 
The bottom panels reveal the presence of complex domains where both phases coexists in a chessboard (also called antiferromagnetic) configuration for large values of the coupling strength. 
\begin{figure}[ht]
\centerline{\includegraphics[width=0.4\textwidth,angle=0]{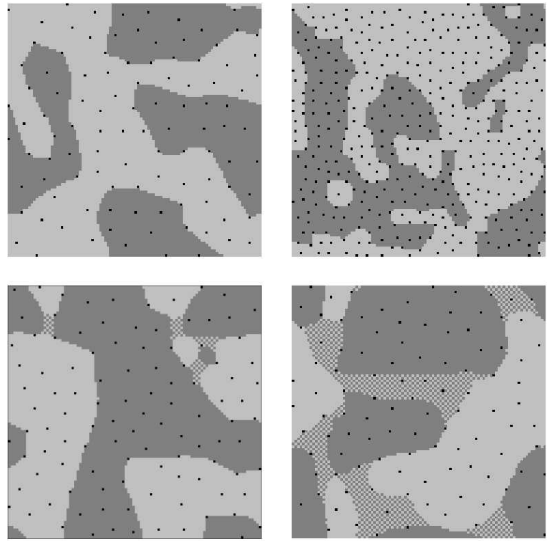}}
\caption{Stationary patterns on a lattice of size $100\times 100$ with impurities, for different values of parameters .
Top left:  $\rho=0.0097 \, (d=8)$, $\epsilon=0.73$. Top right:  $\rho=0.039 \, (d=4)$, $\epsilon=0.73$. Bottom
left: $\rho=0.0097 \, (d=8)$, $\epsilon=0.95$. Bottom right: $\rho=0.0097 \, (d=8)$,  $\epsilon=0.98$.}
\end{figure}

To characterize the phase ordering properties of the system Eq. (\ref{CML}) we use the
normalized size of one phase domain as a function of time
as an order parameter, defined as
\begin{equation}
R_t = \frac{1}{N} \sum_{r=1}^{L/2}\sum_{i,j} 
\delta_{r_{ij},r} \,  \delta_{\sigma^i_t,\sigma^j_t} \, ,
\end{equation}
where $r_{ij}$ is the Euclidean distance between sites $i$ and $j$. Figure~3 shows the average asymptotic value $\langle R_\infty \rangle$ as a function of $\epsilon$ in the absence of impurities. The quantity
$\langle R_\infty \rangle$ exhibits a continuous phase transition from a heterogeneous state, characterized by a small
value of $\langle R_\infty  \rangle$, to a homogeneous state, for which $ \langle R_\infty \rangle =1$, at the critical value $\epsilon_o$ of the coupling parameter reported in \cite{Chate}.
\begin{figure}[ht]
\centerline{\includegraphics[width=0.4\textwidth,angle=90]{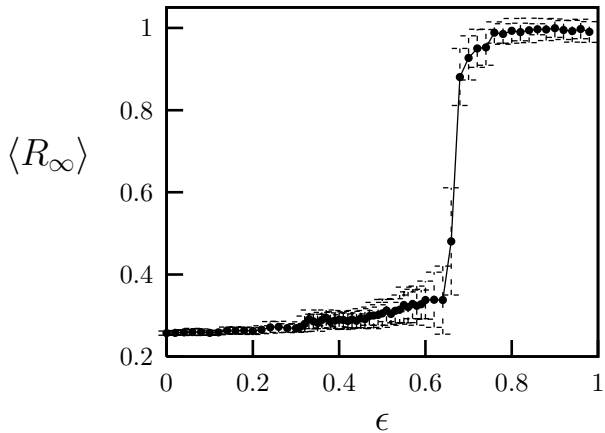}}
\caption{Asymptotic $\langle R \rangle$ as a function of $\epsilon$ for $d=\infty\, (\rho=0)$, averaged over $40$ realizations of initial conditions. Error bars correspond to the standard deviations.}
\end{figure}

Before reaching its stationary value, the average domain size increases in time as $\langle R_t\rangle \sim
t^{\alpha}$, where the exponent $\alpha$ characterizes the rate of phase growth.
Figure~4 shows the evolution of $\langle R_t\rangle$ with fixed value of $\epsilon$ for different values of the distance $d$ between impurities. 
\begin{figure}[b]
\centerline{\includegraphics[width=0.4\textwidth,angle=90]{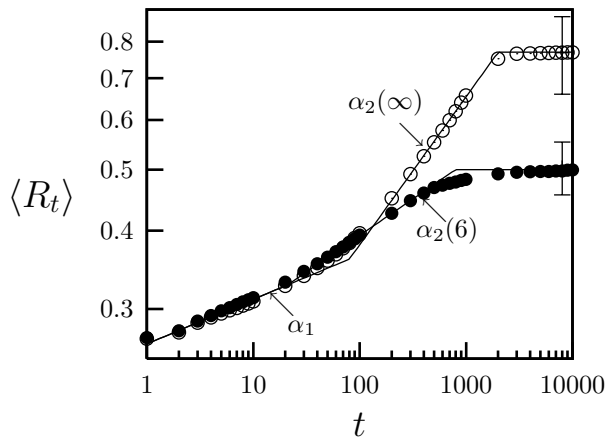}}
\caption{Log-log plot of the average domain size $\langle R_t\rangle$ vs. $t$ for $d=\infty \, (\rho=0)$ (empty circles), and $d=6 \, (\rho=0.017)$ (solid circles), with fixed $\epsilon=0.73$, averaged over $40$ realizations. The values of the scaling exponents $\alpha_1$ and $\alpha_2$, as well as a typical error bar are indicated on each curve.}
\end{figure}
We observe that the phase growth process follows two well differentiated regimes during its time evolution. At early times $t\lesssim 100$, we find $\langle R_t\rangle \sim
t^{\alpha_1}$, with an exponent $\alpha_1 \approx 0.071$ whose value is not appreciably affected by
the presence of impurities. For later times, the scaling behavior changes to $\langle R_t\rangle \sim
t^{\alpha_2}$, with $\alpha_2>\alpha_1$ indicating that domains grow faster in this regime, and where $\alpha_2$ depends on the density of impurities.  Impurities inhibit the growth of the phase domains and prevent the system from reaching a homogeneous state for values of $\epsilon>\epsilon_o$. Figure~4 shows that the average size of the resulting domains in the inhomogeneous state are smaller in the presence of impurities, in agreement with the behavior observed in other
bistable systems on media with impurities \cite{PPR1}. Theoretical models and experiments with binary fluids have also shown that phase growth is restricted when impurities are present \cite{QPGBCJ1,YKAK1} 

The domain growth regimes characterized by the exponents $\alpha_1$ and $\alpha_2$ in Fig.~4 can be understood
in terms of the stability of local configurations of the two phases. With this aim, we define the fraction of sites in a given phase that have $k$ neighbors in that same phase at time $t$, given by 
\begin{equation}
G_t(k) = \frac{1}{N}\sum_{i=1}^N F^i_t(k) \, ,
\end{equation} 
with $k=0,1,2,3,4$; where 
\begin{equation}
F^i_t(k) =\left\lbrace
\begin{array}{ccl}
1, & \mbox{if} & \sum_{j\in {\nu^i}} \delta_{\sigma^i_t,\sigma^j_t} = k\\
\\
0, & \mbox{if} & \sum_{j\in {\nu^i}} \delta_{\sigma^i_t,\sigma^j_t} \neq k \;.
\end{array}\right. 
\end{equation} 

Note that $\sum_k G_t(k)= 1$. 
Figure~5 shows the local spatial configurations corresponding to $k=0,1,2,3,4$. The patterns associated to $k=0,1,2,3$ represent configurations where the two phases are in contact, while the homogeneous state corresponds to $k=4$.
\begin{figure}[ht]
\centerline{\includegraphics[width=0.4\textwidth,angle=0]{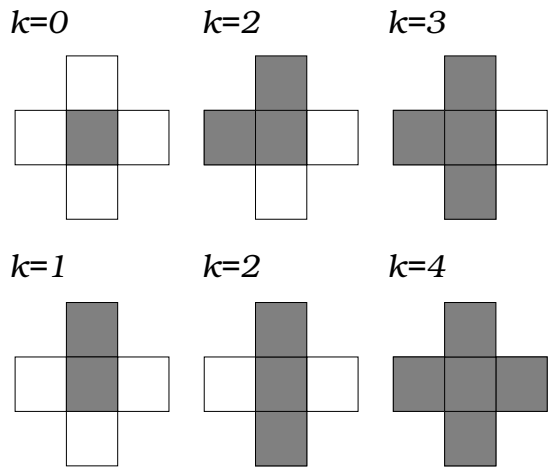}}
\caption{Local configurations taken into account by the fractions $G(k)$, for $k=0,1,2,3$, and $4$.}
\end{figure}

Figure~6 shows the time evolution of the fractions $G_t(k)$ for two different values of $d$. We observe that
in both cases 
\begin{figure}[ht]
\centerline{\includegraphics[width=0.4\textwidth,angle=90]{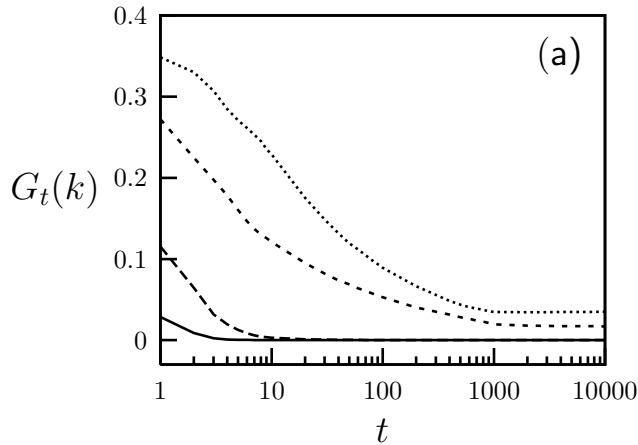}}
\centerline{\includegraphics[width=0.4\textwidth,angle=90]{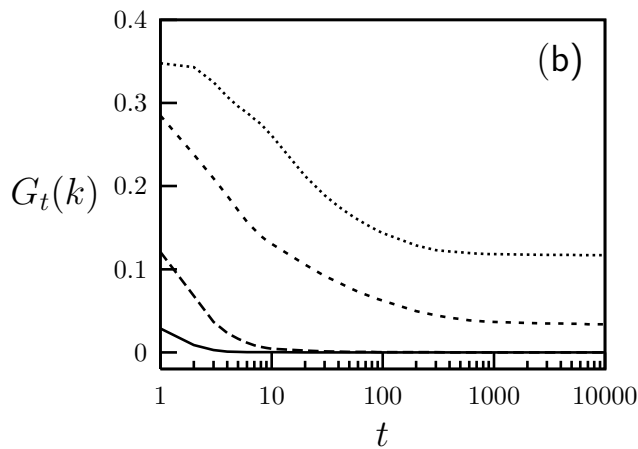}}
\caption{Fractions  $G_t(k)$ vs. $t$ for different values of $d$, with fixed $\epsilon=0.73$. $G_t(0)$ (continuous line); $G_(1)$ (long-dashed line); $G_t(2)$ (short-dashed line); and $G_t(3)$ (dotted line). (a) $d=\infty$.  (b) $d=6$.}
\end{figure}
$G_t(0)$ and $G_t(1)$ vanish at $t\approx 100$. Thus,
the decay of in the number of local configurations associated to $k=0$ and $k=1$ determines the regime of slow
domain growth characterized by the exponent $\alpha_1$ in Fig.~4. 
Phase growth in this earlier regime occurs
by the addition of single cells to domains whose size is of the order of one cell. The spatial scale involved in this
process is smaller than the minimum distance between impurities $d$ and, therefore, the presence of impurities
does not affect this mechanism of phase growth. As a consequence, the exponent $\alpha_1$ does not depend on $d$.

For $t > 100$, the fractions $G_t(2)$ and $G_t(3)$ keep decaying until they reach their asymptotic values.  
Phase growth in this second regime occurs when domains in the same phase and whose average sizes are greater than one cell enter in contact. This process is illustrated in Fig.~7.  
\begin{figure}[ht]
\centerline{\includegraphics[width=0.4\textwidth,angle=0]{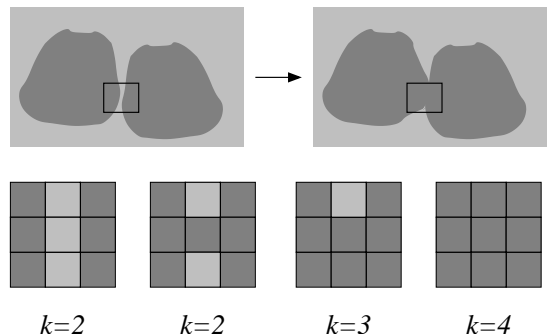}}
\caption{Top: Mechanism of fast phase growth when two domains of the same phase join together to form a larger domain. The arrow indicates the direction of time. Bottom: The local configuration changes occurring in the boxes marked in the top panel are illustrated.}
\end{figure}
At the local level, when two domains of equal phase get separated by a one-cell thick layer of cells in the opposite phase, there are cells on this layer that posses a local configuration described by $k=2$. The coupled map dynamics, Eqs.~(\ref{CML}), acting on a cell in such local configuration
will produce a change of its phase, yielding first another configuration of type $k=2$, then transforming into a configuration of type $k=3$, and finally into the configuration associated to $k=4$. As a result, the domains join together
forming a domain whose average size is much larger than the sizes of the initial domains. Through this growth mechanism, the average size of domains in the system increases faster than by the process the successive additions of single cells that takes place at earlier times.
The difference in the growth velocity of phase domains in these two regimes is manifested by the fact that $\alpha_2 > \alpha_1$.

When impurities are present, the average size of domains is comparable to the minimum distance $d$. Thus, there is a high probability that some impurities lie on the interface. These impurities constrain the transformations of the local configurations $k=2$ into $k=3$, and into $k=4$ described above. Consequently, the phase growth process becomes slower, and the fractions $G_t(2)$ and $G_t(3)$ reach greater asymptotic values in the presence of impurities, as manifested in Fig.~6(b). Therefore, the asymptotic average size of domains for $\rho \neq 0$  is  smaller than that for $\rho=0$,  as it was shown in Fig.~4.

Figure~8 shows the growth exponents $\alpha_1$ and $\alpha_2$ 
as functions of the coupling parameter $\epsilon$, for two different values of the density of impurities $\rho$. We observe that $\alpha_1$ increases slowly with $\epsilon$ and, as it was already manifested in Fig.~4, its behavior is unaffected by the presence of impurities. In contrast, the exponent $\alpha_2$ appears above some threshold value $\epsilon_c >\epsilon_o$
and its behavior depends on both $\epsilon$ and $\rho$. 
\begin{figure}[htb]
\centerline{\includegraphics[width=0.4\textwidth,angle=90]{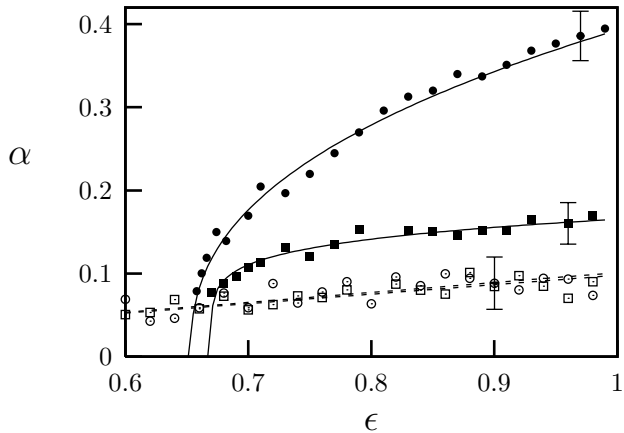}}
\caption{Phase growth exponents $\alpha_1$ (dashed line) and $\alpha_2$ (continuous lines) vs. $\epsilon$ 
for different values of $d$. The values shown are averages obtained over $40$ realizations of initial conditions foe each value of $\epsilon$. Empty circles:
$\alpha_1$ for $d=\infty \, (\rho=0)$; empty squares: $\alpha_1$ for $d=6 \, (\rho=0.017)$; solid squares: $\alpha_2$ for $d=\infty$; solid circles: $\alpha_2$ for $d=6$. Typical error bars are shown on each curve.}
\end{figure}

Figure~9 shows the growth exponent $\alpha_2$ calculated on the space of parameters $(\epsilon, d)$. There is a critical boundary $\epsilon_c(d)$ that separates the regimes where fast growth  occurs, characterized by $\alpha_2>0$, from the regime where only slow growth takes place, corresponding to $\alpha_2=0$, on the plane $(\epsilon, d)$.   Near the critical boundary  $\epsilon_c(d)$, the exponent $\alpha_2$ can be described by the scaling relation $\alpha_2\sim (\epsilon-\epsilon_c)^{\gamma}$, where $\gamma$ also depends on $d$
or,  equivalently, on the density of impurities $\rho$. 
\begin{figure}[b]
\centerline{\includegraphics[width=0.4\textwidth,angle=90]{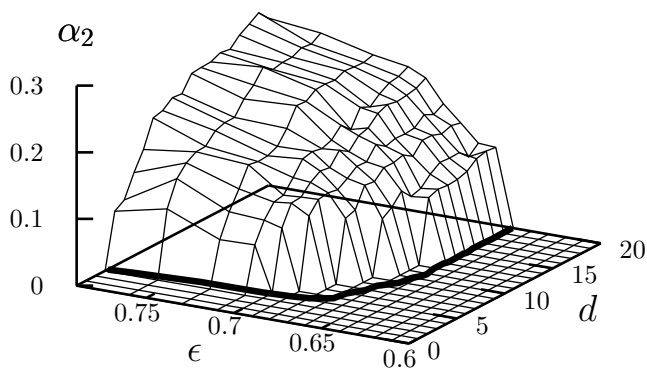}}
\caption{The fast phase growth exponent $\alpha_2$ as a function of $\epsilon$ and $d$. The critical boundary $\epsilon_c(d)$ that separates the fast growth regime from slow growth regime is indicated by a thick continuous line on the plane $(\epsilon,d)$. The values are obtained as in Fig.~8.}
\end{figure}

Figure~10 shows the critical exponent $\gamma$ versus $d$. The exponent $\gamma$ can be well fitted by the scaling relation 
$\gamma(d) \sim (d-d_c)^{\theta}$, where $d_c \approx 4.0$ and $\theta \approx 0.45$. Below  
$d_c \approx 4.0 \, (\rho=0.039)$ there is no regime of fast phase growth.
\begin{figure}[t]
\centerline{\includegraphics[width=0.4\textwidth,angle=90]{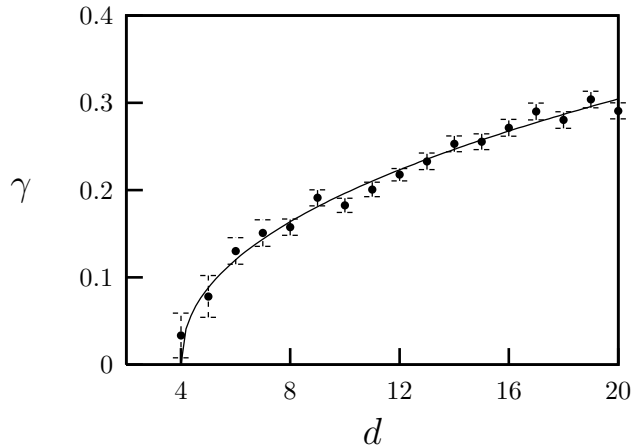}}
\caption{The critical exponent $\gamma$ as a function of the distance between impurities $d$, obtained as averages over $40$ realizations of initial conditions for each parameter value. Error bars are shown.}
\end{figure}

Our results show that it is possible to control both the growth velocity and the size of phase domains in bistable
media by the inclusion of impurities in the spatial support of the system. When the density of impurities increases above some critical value for a given strength of the coupling, both the velocity at which domains grow and their stationary sizes reach smaller values than those corresponding to the absence of impurities. 

To characterize the emergence of chessboard domains for large values of the coupling $\epsilon$, as those shown in the bottom panels of Fig.~2, we calculate the asymptotic fractions $G_\infty(k)$ as functions of $\epsilon$ in Fig.~11. 
\begin{figure}[ht]
\centerline{\includegraphics[width=0.4\textwidth,angle=90]{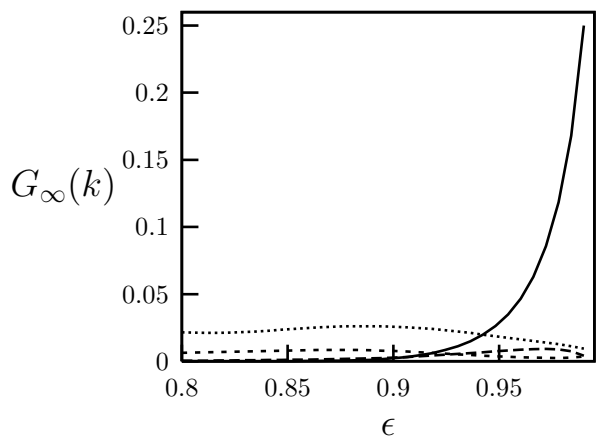}}
\caption{Fractions $G_\infty(k)$ versus $\epsilon$ for $\rho=0$. $G_\infty(0)$ (continuous line); $G_\infty(1)$ (long-dashed line); $G_\infty(2)$ (short-dashed line); and $G_\infty(3)$ (dotted line). Values shown correspond to averages over $40$ realizations of initial conditions.}
\end{figure}
Note that $G_\infty(0)$, which vanished rapidly for small values of $\epsilon$, becomes different from zero and increases for $\epsilon \gtrsim 0.9$. The local configuration corresponding to $k=0$ is effectively associated to a chessboard pattern. Thus the appearance of a finite value of the fraction $G_\infty(0)$ at this large value of $\epsilon$ signals the onset of chessboard domains in the system. Furthermore, we have found that the presence of impurities has little effect on the behavior of $G_\infty(0)$, suggesting that the existence of chessboard domains is mainly associated to large values of the coupling parameter and to the properties of the local dynamics.

Figure~12 summarizes the collective behavior of the coupled map lattice with bistable dynamics in the presence of impurities, Eq.~(\ref{CML}). This figure shows the phase diagram of this system on the space of parameters $(\epsilon,\rho)$. The regions where the homogeneous (one phase) and heterogeneous (two coexisting phases), as well as 
the region where chessboard (antiferromagnetic) states occur are indicated. The critical boundary $\epsilon_c(\rho)$ for the onset of the regime for fast growth of domains $(\alpha_2>0)$ on this plane is marked by a dashed line.
\begin{figure}[ht]
\centerline{\includegraphics[width=0.4\textwidth,angle=90]{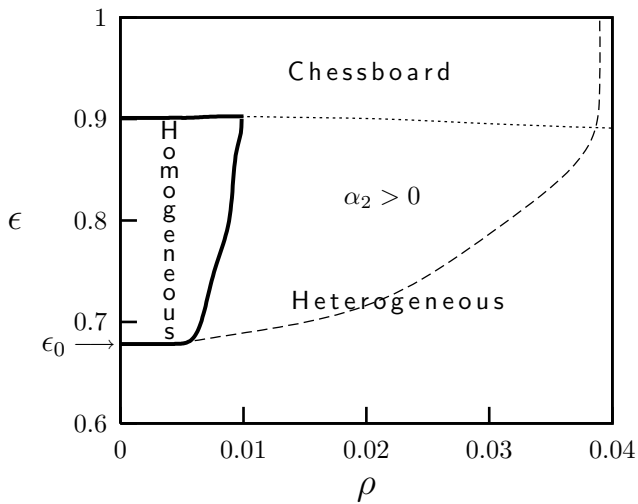}}
\caption{Phase diagram of the system Eq.~(\ref{CML}) on the space of parameters $(\epsilon,\rho)$. The regions where homogeneous, heterogeneous, and chessboard phases exist are indicated by labels. The dashed line indicates the critical boundary for the onset of the regime for fast growth of domains characterized by $\alpha_2 > 0$. The critical value $\epsilon_o$
for the occurrence of domain growth in the absence of impurities is marked on the vertical axis.}
\end{figure}

\section{Conclusions}
We have considered a system of chaotic maps coupled on a lattice with randomly distributed empty sites as a model to study phase ordering processes on media with impurities. The density of impurities $\rho$ is related to the minimum distance between impurities $d$. We have shown that the rate at which phase domains grow becomes smaller when impurities are present in the system. On the other hand, the average size of the resulting domains in the inhomogeneous state of the system decreases when the density of impurities is increased. 

We have calculated the critical boundary that separates the regime of slow growth of domains from the regime of fast growth
on the space of parameters of the system $(\epsilon,d)$. Along this critical boundary, the exponent for fast growth $\alpha_2$ exhibits scaling properties along both parameter axes, manifested by the existence of the critical exponents $\gamma$ and $\theta$ in Figs.~9 and 10. We have explained
the transition between these two growth regimes in terms of the stability properties of the local configuration 
measured by the fractions $G_t(k)$. 

For large values of the coupling strength $\epsilon$, the system displays a complex heterogeneous state consisting of  domains of the two phases coexisting with spatial domains having a chessboard (antiferromagnetic) configuration. We have found that the appearance of chessboard patterns is not appreciably affected by the presence of impurities.
The phase diagram showing the regions where the different behaviors of the system occur on the space of parameters $(\epsilon,\rho)$ was obtained.

Our results indicate that the inclusion of impurities can be used as a control mechanism for the size and growth velocity of domains forming in bistable media. Recent studies of chaotic maps on complex networks \cite{CT2,TCA1} indicate that topology may play a decisive role in determining emerging collective behaviors.
The present results suggest that spatial inhomogeneities may also be employed as a selection mechanism for patterns arising in general spatiotemporal systems.

\section*{ACKNOWLEDGMENTS}
This work was supported in part
by grant I-886-05-02-A from CDCHT,  Universidad de Los Andes, M\'erida, Venezuela.
C.E.  acknowledges support from Decanato de Investigaci\'on, Universidad Nacional Experimental del T\'achira, San Crist\'obal, Venezuela.


\begin{thebibliography}{99}
\bibitem{Kaneko} Chaos 2 (1992) 279, focus issue on
Coupled Map Lattices; edited by K. Kaneko.
\bibitem{Chate} A. Lemaitre and H. Chat\'e, Phys. Rev. Lett. {\bf 82}, 1140  (1999).
\bibitem{Chate2} J. Kockelkoren, A. Lemaitre, and H. Chat\'e, Physica A {\bf 288}, 326 (2000).
\bibitem{Wei} W. Wang, Z. Liu, and B. Hu, Phys. Rev. Lett. {\bf 84}, 2610 (2000).
\bibitem{Stra} L. Angelini, M. Pellicoro, and S. Stramaglia, Phys. Lett. A {\bf 285},
293 (2001).
\bibitem{Just}  F. Schm\"user, W. Just, and H. Kantz, Phys. Rev. E {\bf 61}, 3675 (2000).
\bibitem{A1}  L. Angelini, Phys. Lett. A {\bf 307}, 41 (2003).
\bibitem{TCA1} K. Tucci, M. G. Cosenza and O. Alvarez-Llamoza, Phys. Rev. E 
{\bf 68}, 027202 (2003).
\bibitem{MH1} J. Miller and D. Huse, Phys. Rev. E {\bf 48}, 2528 (1993).
\bibitem{PPR1} R. Paul, S. Puri and H. Rieger, Phys. Rev. E {\bf 71}, 61109 (2005).
\bibitem{YKAK1} K. Yurekli, A. Karim, E. J. Amis y R. Krishnamoorti, Macromolecules {\bf 36}, 7256 (2003).
\bibitem{QPGBCJ1} F. Qiu, G. Peng, V. V. Ginzburg, A. C. Balazs, H. Y. Chen, D. Jasnow, J. Chem. Phys. {\bf 115}, 3779 (2001).
\bibitem{CT2} M. G. Cosenza and K. Tucci, Phys. Rev. E {\bf 65}, 036223 (2002).
\end{thebibliography}
\end{document}